\documentclass[final, journal, twocolumn]{IEEEtran}
\usepackage[utf8]{inputenc}
\usepackage[english]{babel}
\usepackage{graphicx,color}
\usepackage{amssymb,amsmath,amsfonts}
\usepackage{cite}
\usepackage{flushend}
\usepackage{float}
\usepackage{amsmath}
\usepackage{siunitx}
\usepackage{physics}
\begin{document}

  \title{Coupling of Magneto-Thermal and Mechanical Superconducting Magnet Models by Means of Mesh‑Based Interpolation}

\author{{M. Maciejewski, P. Bayrasy, K. Wolf, M. Wilczek, B. Auchmann, T. Griesemer, L. Bortot, M. Prioli, A.M. Fernandez Navarro, S. Sch\"ops, I. Cortes Garcia, and A.P. Verweij}

\thanks{
M. Maciejewski is with CERN, Switzerland, and Institute of Automatic Control, Technical University of Łódź, 18/22 Stefanowskiego St., Poland (\mbox{e-mail}: michal.maciejewski@cern.ch).

P. Bayrasy and K. Wolf are with Fraunhofer Institute SCAI, Schloss Birlinghoven, 53757 Sankt Augustin, Germany.

M. Wilczek is with Technical University of Łódź, 18/22 Stefanowskiego St., Poland.

B.Auchmann is with CERN, Switzerland, and with Paul Scherrer Institute, 5232 Villigen PSI, Switzerland.

S. Sch\"ops and I. Cortes Garcia are with Technische Universit\"at Darmstadt, Karolinenpl. 5, 64289 Darmstadt, Germany.

L. Bortot, M. Prioli and A.P. Verweij are with CERN, Switzerland.

T. Griesemer is with RWTH Aachen, Templergraben 55, 52056 Aachen.}}

\maketitle
\selectlanguage{english}
\begin{abstract}
In this paper we present an algorithm for the coupling of magneto-thermal and mechanical finite element models representing superconducting accelerator magnets. The mechanical models are used during the design of the mechanical structure as well as the optimization of the magnetic field quality under nominal conditions. The magneto-thermal models allow for the analysis of transient phenomena occurring during quench initiation, propagation, and protection. Mechanical analysis of quenching magnets is of high importance considering the design of new protection systems and the study of new superconductor types. 
We use field/circuit coupling to determine temperature and electromagnetic force evolution during the magnet discharge. These quantities are provided as a load to existing mechanical models. 
The models are discretized with different meshes and, therefore, we employ a mesh-based interpolation method to exchange coupled quantities.  The coupling algorithm is illustrated with a simulation of a mechanical response of a standalone high-field dipole magnet protected with CLIQ (Coupling-Loss Induced Quench) technology.
\end{abstract}


\IEEEpeerreviewmaketitle

\section{Introduction} \label{Intro}

Superconducting high-field magnets are one of the main components of high-energy particle accelerators. Due to the considerable energy stored in the magnetic field, a transition of a small cable volume from the normal conducting to the superconducting state, also known as a quench, may result in an uncontrolled release of the energy as ohmic loss and possibly in damage of magnet or circuit components. 

The Finite Element Method (FEM) is used to perform accurate modeling of superconducting magnets for magnetic field design, mechanical design, and the calculation of the peak temperature and voltages to ground in the magnet during a quench. 
In each design step a set of partial differential equations with appropriate boundary conditions, nonlinear material properties, and problem-adapted meshes is solved. For as long as possible, the physical problems are treated independently and solved with dedicated FEM programs. The study of multi-physics phenomena, however, requires to solve coupled systems of equations, and monolithic approaches by a single tool might not be desirable or available.  

Co-simulation treats the coupled models independently by means of input and output relations. The Simulation of Transient Effects in Accelerator Magnets (STEAM) project at CERN follows this approach and incorporates the waveform relaxation method to perform the coupling of magneto-thermal field models with electrical circuits \cite{Cortes-Garcia_2017ab}, \cite{STEAM2017}. The main objective of this paper is to design an algorithm to couple magneto-thermal and mechanical models.  Thus, we extend the framework by introducing an additional one-way coupling of magneto-thermal transient analysis to a static mechanical response of a magnet subject to temperature and Lorentz force variation. In this setting, the interpolation of the results obtained with two different meshes has to be addressed. 

Such a mechanical analysis is especially important in case of superconducting coils made of a brittle material, e.g. $\mathrm{Nb}_3\mathrm{Sn}$. The application of a CLIQ, (Coupling-Loss Induced Quench) \cite{ravaioli2014new}, \cite{ravaioli2015cliq} system to protect these coils poses a question on the mechanical impact of the current overshoot during the magnet discharge. The magneto-thermal study is performed with COMSOL Multiphysics \cite{COMSOL} and the mechanical analysis with ANSYS APDL \cite{ANSYS}. We use the MpCCI (Multi-physics Code Coupling Interface) \cite{MpCCI} environment, which provides a generic coupling mechanism (in particular mesh-based interpolation) and was used to solve coupled problems in other fields, e.g. \cite{Schreiber2006}, \cite{Bayrasy2016}.

An analysis of the Lorentz force impact on the superconducting magnet structure has been already studied. In \cite{Milanese2010} a method for the interpolation of concentrated electromagnetic forces obtained with the electromagnetic FEM-BEM code ROXIE \cite{ROXIE} to a mechanical ANSYS model has been proposed. The authors of \cite{Caspi2006} proposed an integrated approach to thermal, electrical, and structural analysis of the magnet design process involving several CAD and simulation tools.

The rest of the paper is organized as follows. Section~II introduces the models. The one-way coupling algorithm is discussed in Section~III. In Section~IV we present a simulation of an 11 T dipole magnet for the High-Luminosity upgrade of the Large Hadron Collider \cite{Nilsson2017}. In the study, the magnet is protected with the CLIQ technology.

\section{Mathematical Models} \label{models}
A mechanical analysis of a superconducting magnet replicates the magnet-assembly operations and it is used to find proper dimensions and materials for the structural elements. As a result, after assembling and cooling down, the elements are in good contact and after powering the magnet turns remain in compression. We extend this analysis by studying a current discharge and what is the impact of the resulting Lorentz force and temperature evolution on the magnet structure. 

\subsection{Magneto-Thermal Model}\label{magneto_thermal}
A detailed derivation of the 2D finite element magneto-thermal model can be found in \cite{Bortot2017}. The model takes into account inter-filament and inter-strand coupling currents \cite{wilson1983superconducting} occurring in superconducting Rutherford cables through equivalent magnetization \cite{deGeresem2004finite}, saturation of the iron yoke as well as nonlinear material properties in the thermal model. The fundamental entity representing a coil in the magneto-thermal mode is a half-turn over which material properties and physical laws are homogenized. Hereunder we report the equations governing the model.

Considering a magnetoquasistatic setting ($\partial_t\vec{D}=0$ with $\vec{D}$ being the electric flux density) and representing the magnetic flux density $\vec{B}$ in terms of the magnetic vector potential $\vec{A}$ as $\vec{B}=\nabla \times \vec{A}$, the magnetic problem is governed by the following partial differential equation

\begin{equation}
 \nabla \times (\nu \nabla \times \vec{A}) + \nabla \times \left(\nu \tau _{\mathrm{eq}}\nabla \times \partial_t \vec{A}\right) = \vec{\chi } I ,
\label{eq:curl_curl}
\end{equation}
with appropriate boundary conditions depending on the model symmetry. Here, $\nu$ is the magnetic permeability, $\tau_\mathrm{eq}$ is the equivalent time constant of cable eddy currents as given in \cite{verweij1995electrodynamics}, $\vec{\chi}$ is the winding density matrix, and $I$ is the current flowing through the coil which is determined by the field/circuit coupling \cite{Cortes-Garcia_2017ab}.

The temperature ${T}$ in each coil half-turn is determined from the heat balance equation with adiabatic boundary conditions
\begin{equation}
{C_{p}}{\partial_t }{T} = -\triangledown\cdot\vec{q} + Q_\mathrm{ohmic} + {Q_\mathrm{M}},
\label{eq:heat_balance}
\end{equation}
where ${C_{p}}$ is the volumetric heat capacity of the coil domain, the heat conduction $\triangledown\cdot\vec{q}$ is defined by Fourier's Law, $Q_\mathrm{ohmic}$ and ${Q_\mathrm{M}}$ are ohmic loss and losses due to cable coupling currents in the half-turns per unit volume, respectively.

Temperature distribution in the coil $T$ and the Lorentz force $\vec{F}_\mathrm{L}=\vec{J} \times (\nabla \times \vec{A})$, with $\vec{J}=\vec{\chi} I$ being the current density, are obtained from the field solution and act as a load for the mechanical problem. 

\subsection{Mechanical Model}\label{mechanical}

The mechanical model is represented as a static linear elastic equation, the so-called Navier-Cauchy equation\cite{logan2012}
\begin{equation}
\mu \Delta \vec{u} + (\lambda + \mu) \nabla (\nabla \cdot \vec{u})+\vec{F}_T(T)+\vec{F}_\mathrm{L}(\vec{J},\vec{A})=0
\label{eq:linear_elasticity}
\end{equation}
where $\lambda$ and $\mu$ are the Lam\'e parameters, $u$ the displacement vector, and the thermal force $F_T(T)$ is given by
\begin{equation}
\vec{F}_T(T) = \nabla \cdot C \alpha \Delta T,
\label{eq:thermal_force}
\end{equation}
where $C$ is the elasticity matrix, and $\alpha$ is the thermal expansion coefficient. Stress and strain are derived from the field solution and allow for a quantification of mechanical load on the coil. 

Both models, (\ref{eq:curl_curl}-\ref{eq:heat_balance}) and (\ref{eq:linear_elasticity}) are discretized by means of a low order finite element discretization on a quadrilateral and/or triangular grid. 

\section{Mesh-based Interpolation} \label{sec:mesh_based_interpolation}
Mesh-based interpolation considers two or more coupled models with different mesh definitions. Mesh definitions consist of a vector of mesh nodal positions and a connectivity matrix providing the relationship between mesh elements and mesh nodes. In order to exchange quantities obtained with these models, solutions located at mesh entities have to be accurately interpolated based on the mesh definitions and element interpolation functions of the respective models. The MpCCI coupling environment already supports ANSYS models and automatically reads their mesh definition. For COMSOL it was necessary to develop a dedicated Java code adapter based on the MpCCI API (Application Programming Interface). The COMSOL code adapter is capable of reading mesh definitions and quantities at the mesh nodes.

\begin{figure}[H]
	\centering
	\includegraphics[width=0.5\textwidth]{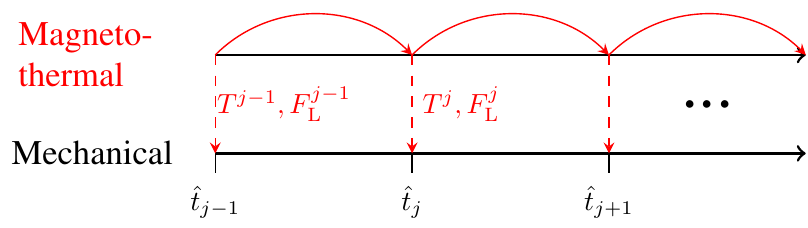}
	\caption{\label{fig:one_way_coupling} Scheme of the one-way information exchange between magneto-thermal (red) and mechanical (black) models.}
\end{figure}

For the purpose of a one-way coupling, the overall simulation time $\mathcal{I}$ is divided into $N$ time windows $\mathcal{I}_j = (\hat{t}_j, \hat{t}_{j+1}]$ with $j=0,...,N-1$. The magneto-thermal model is assumed to be already solved for $t \in (\hat{t}_0, \hat{t}_{N}]$ before executing the algorithm. The computed field solution retrieved at discrete time points $\hat{t}_j$ becomes a load for a static mechanical analysis. The mechanical model is initially loaded with pre-stress and cool-down steps and then solved at the communication points $\hat{t}_j$. In case of the considered coupled problem, with $j$ denoting the time window index, the algorithm takes the following steps as depicted in Fig. \ref{fig:one_way_coupling}:

\begin{enumerate}
\item[0)] Set $j=0$ and initialize the models, define the coupling region and provide its mesh definition. The magneto-thermal model (\ref{eq:curl_curl}-\ref{eq:heat_balance}) is solved with an external current profile $I$ and  the mechanical model performs initial pre-stress and cool-down calculations.

\item[1)] Read $T^j$ and $F_\mathrm{L}^j$ for $t=\hat{t}_j$ from the COMSOL model nodes and send them to the MpCCI server.

\item[2)] Perform a mesh-based interpolation on the MpCCI server. 

\item[3)] Receive $T^j$ and $F_\mathrm{L}^j$ and solve a static analysis (\ref{eq:linear_elasticity}) in ANSYS.

\item[4)] If $j=N$ terminate the co-simulation, otherwise set $j=j+1$ and go to point 1) to start the next time window.
\end{enumerate}

The algorithm is realized as two independent time loops interacting with the MpCCI server. The ANSYS and COMSOL time loops are implemented, respectively, as APDL and Java codes. Four commands are necessary for the interaction with the MpCCI server, i.e. server initialization, synchronization of model execution, transfer of coupled quantities, and termination of the connection to the MpCCI server. 

\section{Numerical Example}
\begin{figure}
	\centering
		\includegraphics[width=0.35\textwidth]{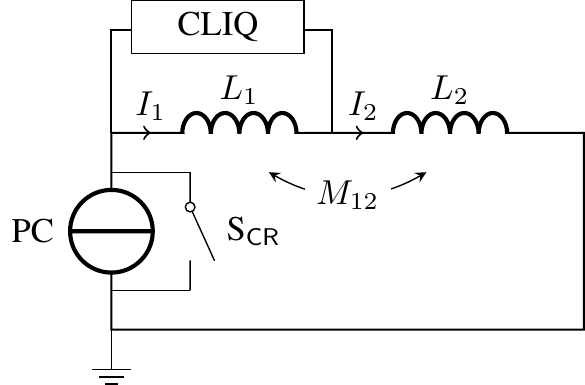}
	\caption{Schematic of the standalone circuit with an 11 T dipole magnet powered by a power converter (PC) with a crowbar (S$\mathsf{_{CR}}$). Magnet is protected by a CLIQ system modelled as a charged capacitor bank and a thyristor triggered at $t=0~\si{\second}$.}
	\label{fig:circuit_standalone}
\end{figure}

One-way coupling of the electro-thermal and mechanical model with mesh-based interpolation is illustrated by means of a case study of a standalone, single-aperture, 5.5-meter long, 11 T dipole magnet protected by a CLIQ system as depicted in Fig. \ref{fig:circuit_standalone}. Parameters of the magnet can be found in \cite{Nilsson2017}. The initial current of the power converter (PC) is equal to $11850~\si{\ampere}$ and is switched off at $t=0~\si{\second}$, which is also the triggering time of a CLIQ unit. The CLIQ unit includes a capacitor bank of capacitance $60~\si{\milli\farad}$ charged to an initial voltage of $500~\si{\volt}$. The equivalent circuital representation of the magnet is composed of two self inductances 
$L_1=L_2=14~\si{\milli\henry}$ of upper and lower pole, respectively, and a mutual inductance $M_{12}=M_{21}=7.1~\si{\milli\henry}$.

\begin{figure}
	\centering
	\includegraphics[width=0.45\textwidth]{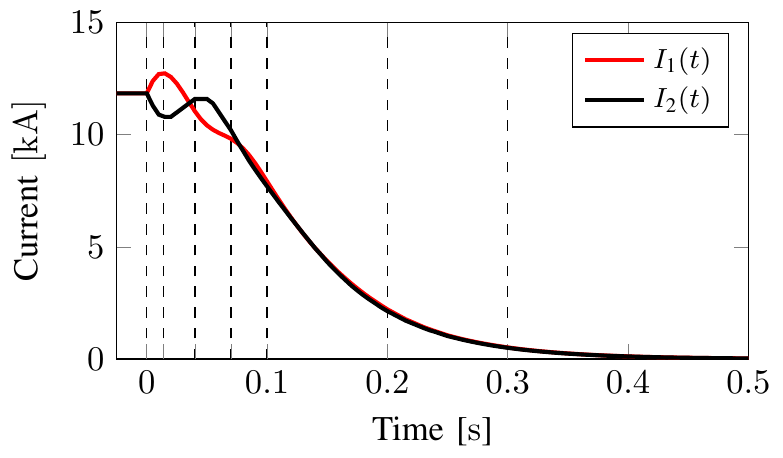}
	\caption{Time evolution of the current discharge in two poles of a magnet following the CLIQ triggering. Dashed lines indicate time points for which results of mechanical analysis are reported.}
	\label{fig:cliq_current}
\end{figure}

The current profile calculated with a field/circuit coupling algorithm based on waveform relaxation scheme \cite{Cortes-Garcia_2017ab} is shown in Fig. \ref{fig:cliq_current}. The algorithm exchanges coupled variables in form of waveforms and solves both magneto-thermal and circuit models separately over a period of time. Once the time window is executed, another exchange takes place until the convergence is reached \cite{Cortes-Garcia_2017ab}. The electrical network is modeled with ORCAD PSpice and the magneto-thermal coupled problem (\ref{eq:curl_curl}-\ref{eq:heat_balance}) is solved in COMSOL Multiphysics \cite{Bortot2017}. The co-simulation time is equal to $\hat{t}_{N}=500~\si{\milli\second}$. In order to study thermal stress due to the initial quench, we assume the high-field inner layer turn to be artificially brought to the normal conducting state $25~\si{\milli\second}$ prior to the CLIQ unit triggering at $t=0~\si{\milli\second}$. In case of a long-enough magnet (as a rule of thumb, a magnet with ratio of the magnetic length to the aperture size greater than 20) a 2D model serves as a good approximation. Due to the presence of a current imbalance in the upper and lower coils and taking into account the model symmetry, the model is composed of two quadrants. The coupling region of each model is limited to the coil domain.  

For the solution of (\ref{eq:linear_elasticity}) we consider a  mechanical model implemented in ANSYS APDL \cite{Savary2015}. The model contains all the main structural components of a magnet, i.e. outer shell, iron yoke, steel collar, copper wedges and the coil with  turns composed of $\mathrm{Nb}_3\mathrm{Sn}$, $\mathrm{Cu}$, insulation, and resin. There are appropriate contact elements for connections between the structural components.

\begin{figure}
\begin{center}
  \includegraphics[height=52mm]{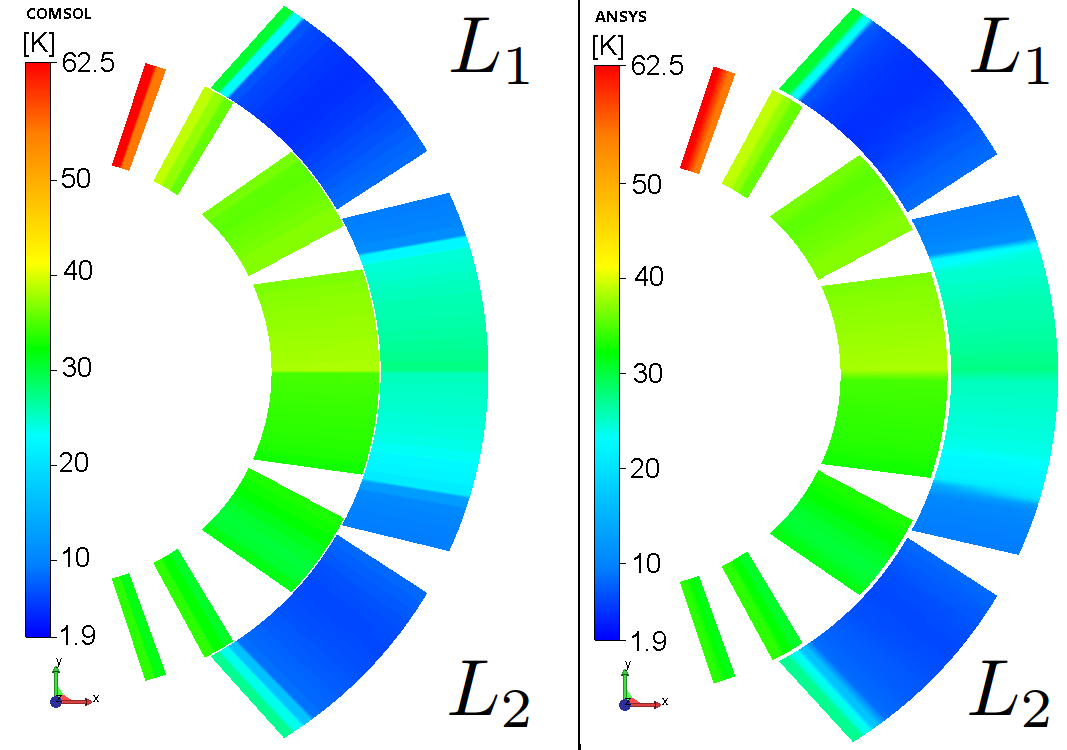}
  \caption{Temperature comparison at $t=14~\si{\milli\second}$ between COMSOL (left) and ANSYS (right) models.}
  \label{fig:temperature_comparison}
\end{center}
\vspace{-1em}
\end{figure}

\begin{figure}
\begin{center}
  \includegraphics[height=52mm]{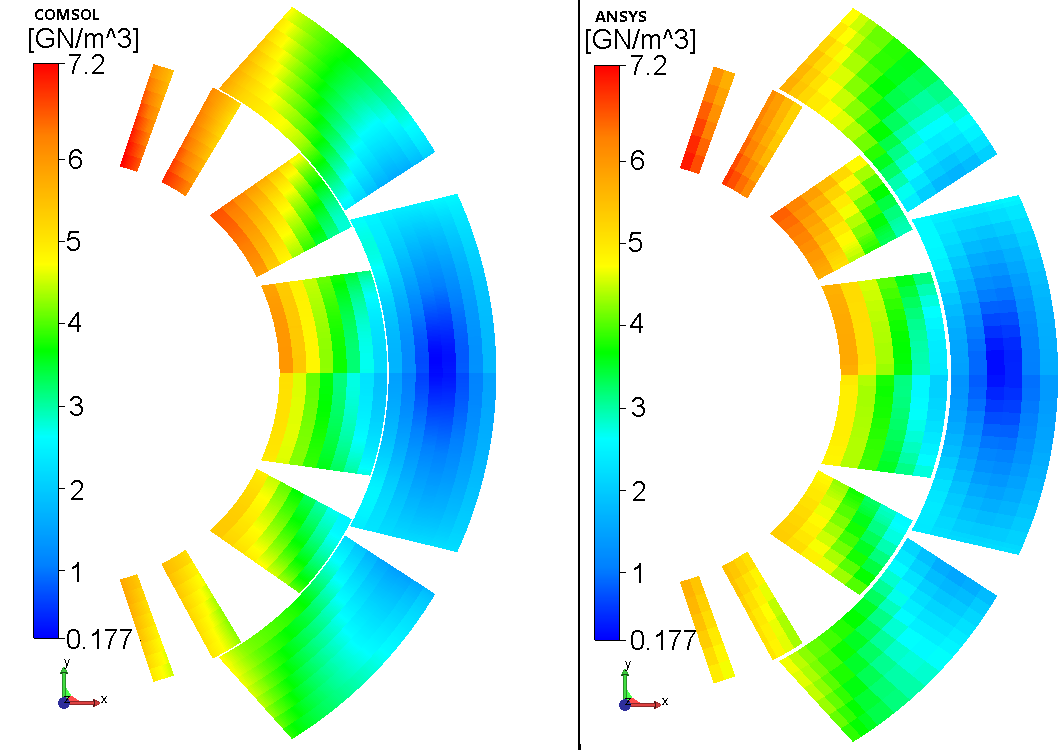}
  \caption{Comparison of the Lorentz force transfer at $t=14~\si{\milli\second}$ between COMSOL (left) and ANSYS (right) models.}
  \label{fig:lorentz_force_comparison}
\end{center}
\vspace{-1em}
\end{figure}

The algorithm described in Section \ref{sec:mesh_based_interpolation} was executed for $N=103$ time points. In order to properly resolve the initial oscillation, for $t \in [0~\si{\milli\second},30~\si{\milli\second})$ field solution from the magneto-thermal model is retrieved every $1~\si{\milli\second}$ time step, whereas for $t \in [30~\si{\milli\second},500~\si{\milli\second}]$, every $10~\si{\milli\second}$.Figures \ref{fig:temperature_comparison} and \ref{fig:lorentz_force_comparison} present temperature and Lorentz Force distributions, respectively, in both models at $t=14~\si{\milli\second}$ for which the maximum CLIQ current is observed. The temperature is extracted from the COMSOL element nodes, interpolated by the MpCCI server and assigned to nodes of the mechanical model mesh elements. The Lorentz force density is evaluated at the element centers on the COMSOL side and interpolated onto centroids of the mesh elements used to discretize the ANSYS model. Next, the interpolated force is applied as a load on the nodes of the element. CLIQ oscillations introduce a variation of the magnetic field which causes heat deposition due to the inter-filament and inter-strand coupling losses in the coil. As a result large fractions of the coil volume transition to the normal conducting state, for which the temperature is increased further by the ohmic loss deposition. The CLIQ-induced current imbalance is demonstrated as an asymmetry of the Lorentz force distribution between the upper and the lower pole. Effectively the electrodynamic force superimposes with the thermal stress in the magnet cross-section.

\begin{figure}
\begin{center}
  \includegraphics[height=52mm]{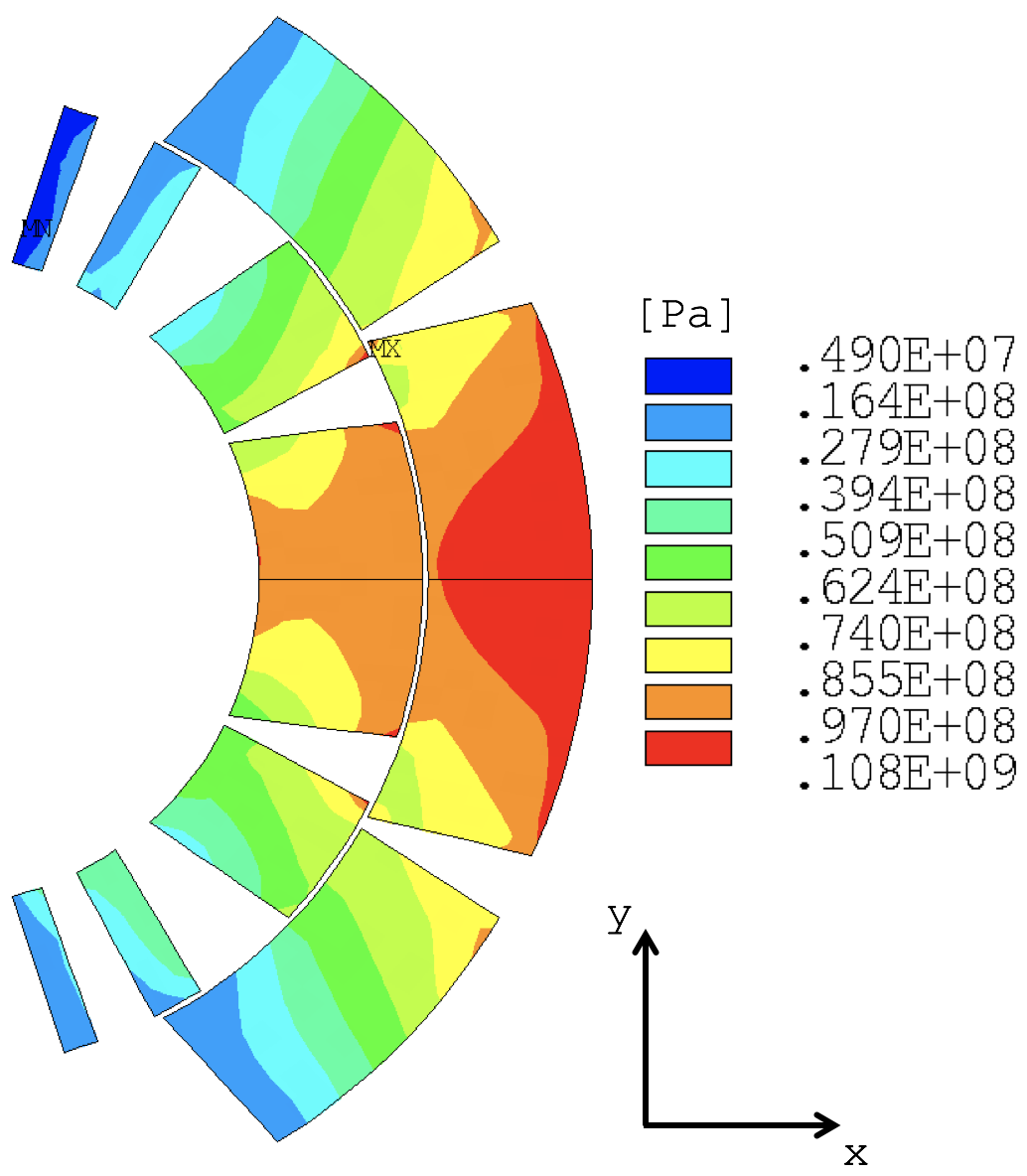}
  \caption{Equivalent stress in the coil at $t=14~\si{\milli\second}$.}
  \label{fig:equivalent_stress}
\end{center}
\vspace{-1em}
\end{figure}

The maximum stress corresponding to the peak current ($t=14~\si{\milli\second}$) is reported in Fig. \ref{fig:equivalent_stress}.
The peak value is equal to $108~\si{\mega\pascal}$ and is located in the bottom-left corner of the second inner layer block of the upper pole (note that the mid-plane blocks are numbered as the first).
Table \ref{tab:stress_vs_time} provides a summary of maximum stress in the coil domain for the time points indicated in Fig. \ref{fig:cliq_current}. The first time points correspond to the initial CLIQ-induced oscillations, when the Lorentz force reaches the largest value in the upper pole. Additionally, time points for  the middle and the final phase of the discharge are considered when the temperature gradients become significant.

\begin{table}[H]
\caption{\label{tab:stress_vs_time} Summary of Maximum Equivalent Stress of the Mechanical Model at Selected Time Points.}
\centering
\begin{tabular}{ccc}
$t~[\si{\milli\second}]$ & 
$\mathrm{max}(\sigma_{\mathrm{eq}})~[\si{\mega\pascal}]$ &
Location\\
\hline
-25 & 102 & second block of inner layer (upper pole)\\
0 & 102 & second block of inner layer (lower pole)\\
14 & 108 & second block of inner layer (upper pole)\\
40 & 95.7 & first block of inner layer (upper pole) \\
70 & 85.6 & first block of inner layer (lower pole) \\
100 & 93.1 & third block of inner layer (upper pole)\\
200 & 122 & third block of inner layer (upper pole)\\
300 & 124 & third block of inner layer (upper pole)\\
\end{tabular}
\end{table}

During the initial period of the discharge the maximum stress is mainly due to the Lorentz force. Furthermore, due to the CLIQ oscillations the mechanical structure is subject to local increase of the Lorentz force beyond the nominal value. Simultaneously, thermal stresses are introduced since the CLIQ system is capable of quenching large portions of the coil volume in a short time. As the current decreases, the electromagnetic force is reduced below the nominal value and at the same time the temperature gradients start to be dominating. Indeed, at $t=100~\si{\milli\second}$ the maximum equivalent stress is increasing again due to the thermal stress. For the last considered time point the maximum equivalent stress occurs at the block adjacent to the initial quench zone. The hot-spot temperature is equal to 145 K and is also located in that block.

\section{Conclusion}

In this paper, a mesh-based interpolation method for co-simulation of magneto-thermal and mechanical finite element models has been presented. The one-way coupling algorithm is used to study the mechanical impact of temperature gradients and Lorentz force evolution during a quench and subsequent circuit-energy discharge in a superconducting magnet. In order to exchange variables defined over non-conforming mesh definitions we employ MpCCI.

The algorithm has been illustrated by means of a case study of the 11 T dipole magnet in a standalone setting protected by a CLIQ system. The magneto-thermal field model was solved with a current profile obtained as a solution of a field/circuit coupling algorithm. Additionally, a high-field turn was assumed to be in the normal conducting state prior to the CLIQ trigger in order to mimic the initial quench development. The mechanical model was initialized with pre-stress and cool-down studies. Then, we carried out an analysis of the results obtained from the coupling of magneto-thermal and mechanical models during the CLIQ operation.

To conclude, we note that the superimposed effect of the electrodynamic forces and temperature gradients either amplify or compensate one another in certain parts of a coil depending on the considered scenario, i.e. the initial quench location, CLIQ configuration, magnet geometry. Therefore, the analysis of the mechanical response of the magnet structure during the quench protection is important in order to estimate the magnitude and location of the peak stress.
Future research may exploit the MpCCI environment feature of bi-directional coupling schemes, which could be used to account for the strain dependence of the superconducting critical surface during a quench.

\section{Acknowledgments}
Authors would like to thank Matthias Mentink, Christian L\"offler, and Etienne Rochepault from CERN for consultations on mechanical ANSYS APDL models. This work has been partially supported by the Excellence Initiative of the German Federal and State Governments and the Graduate School of CE at TU Darmstadt.


\begin{thebibliography}{10}
\providecommand{\url}[1]{#1}
\csname url@samestyle\endcsname
\providecommand{\newblock}{\relax}
\providecommand{\bibinfo}[2]{#2}
\providecommand{\BIBentrySTDinterwordspacing}{\spaceskip=0pt\relax}
\providecommand{\BIBentryALTinterwordstretchfactor}{4}
\providecommand{\BIBentryALTinterwordspacing}{\spaceskip=\fontdimen2\font plus
\BIBentryALTinterwordstretchfactor\fontdimen3\font minus
  \fontdimen4\font\relax}
\providecommand{\BIBforeignlanguage}[2]{{
\expandafter\ifx\csname l@#1\endcsname\relax
\typeout{** WARNING: IEEEtran.bst: No hyphenation pattern has been}
\typeout{** loaded for the language `#1'. Using the pattern for}
\typeout{** the default language instead.}
\else
\language=\csname l@#1\endcsname
\fi
#2}}
\providecommand{\BIBdecl}{\relax}
\BIBdecl

\bibitem{Cortes-Garcia_2017ab}
I.~Cortes~Garcia, S.~Schöps, M.~Maciejewski, L.~Bortot, M.~Prioli, B.~Auchmann,
  and A.~Verweij, ``Optimized field/circuit coupling for the simulation of
  quenches in superconducting magnets,'' \emph{IEEE Journal on Multiscale and
  Multiphysics Computational Techniques}, vol.~2, no.~1, pp. 97--104, May 2017.

\bibitem{STEAM2017}
L.~Bortot, B.~Auchmann, I.~Cortes~Garcia, A.~Fernandez~Navarro, M.~Maciejewski,
  M.~Mentink, M.~Prioli, S.~Schöps, E.~Ravaioli, and A.~Verweij, ``STEAM: A
  hierarchical co-simulation framework for superconducting accelerator magnet
  circuits,'' \emph{IEEE Transactions on Applied Superconductivity}, vol.~28, no.~3, 
  Apr. 2018.

\bibitem{ravaioli2014new}
E.~Ravaioli, V.~Datskov, C.~Giloux, G.~Kirby, H.~ten Kate, and A.~Verweij,
  ``New, coupling loss induced, quench protection system for superconducting
  accelerator magnets,'' \emph{IEEE Transactions on Applied Superconductivity},
  vol.~24, no.~3, pp. 1--5, June 2014.

\bibitem{ravaioli2015cliq}
E.~Ravaioli, ``{CLIQ}. {A} new quench protection technology for superconducting
  magnets,'' Ph.D. dissertation, Universiteit Twente, 2015.

\bibitem{COMSOL}
\BIBentryALTinterwordspacing
{\relax COMSOL Multiphysics} 5.2a. [Online]. Available: \url{http://comsol.com}
\BIBentrySTDinterwordspacing

\bibitem{ANSYS}
\BIBentryALTinterwordspacing
{\relax ANSYS APDL} 16.2. [Online]. Available: \url{http://ansys.com}
\BIBentrySTDinterwordspacing

\bibitem{MpCCI}
\BIBentryALTinterwordspacing
{\relax Fraunhofer Institute}.~SCAI. {\relax MpCCI Coupling Environment 4.5}.
  [Online]. Available: \url{https://www.mpcci.de}
\BIBentrySTDinterwordspacing

\bibitem{Schreiber2006}
U.~Schreiber and U.~van Rienen, ``\BIBforeignlanguage{english}{Coupled
  calculation of electromagnetic fields and mechanical deformation},'' in
  \emph{\BIBforeignlanguage{english}{Scientific Computing in Electrical
  Engineering {SCEE} 2004}}, ser. Mathematics in Industry, A.~M. Anile,
  G.~Alì, and G.~Mascali, Eds., no.~9., Berlin, Germany: Springer, 2006.

\bibitem{Bayrasy2016}
P.~Bayrasy, ``{\relax Coupled simulations of electric arcs for switching
  devices with MpCCI and ANSYS},'' in \emph{NAFEMS European Conference
  "Multiphysics Simulation" 2016}.\hskip 1em plus 0.5em minus 0.4em\relax IEEE,
  2016, pp. 39--41.

\bibitem{Milanese2010}
A.~Milanese, ``A method to transfer concentrated Lorentz forces to a finite
  element mechanical model,'' \emph{EuCARD Publication}, pp. 667--670, 2010.

\bibitem{ROXIE}
\BIBentryALTinterwordspacing
S.~Russenschuck. {\relax ROXIE} 10.2. [Online]. Available:
  \url{http://cern.ch/roxie}
\BIBentrySTDinterwordspacing

\bibitem{Caspi2006}
S.~Caspi and P.~Ferracin, ``Toward integrated design and modeling of high field
  accelerator magnets,'' \emph{IEEE Transactions on Applied Superconductivity},
  vol.~16, no.~2, pp. 1298--1303, June 2006.

\bibitem{Nilsson2017}
E.~Nilsson, S.~I. Bermudez, A.~Ballarino, B.~Bordini, L.~Bottura, J.~Fleiter,
  F.~Lackner, C.~Loffler, J.~C. Perez, H.~Prin, G.~DeRijk, D.~Smekens, and
  F.~Savary, ``{\relax Design Optimization of the Nb3Sn 11 T Dipole for the
  High Luminosity LHC},'' \emph{IEEE Transactions on Applied
  Superconductivity}, vol.~27, no.~4, June 2017.

\bibitem{Bortot2017}
L.~Bortot, B.~Auchmann, I.~Cortes~Garcia, A.~Fernandez~Navarro, M.~Maciejewski,
  M.~Prioli, S.~Schöps, and A.~Verweij, ``A 2-d finite-element model for
  electro-thermal transients in accelerator magnets,'' \emph{IEEE Transactions on 
	Magnetics}, vol.~54, no.~3, Mar. 2018.

\bibitem{wilson1983superconducting}
M.~N. Wilson, \emph{Superconducting magnets}.\hskip 1em plus 0.5em minus
  0.4em\relax Clarendon Press Oxford, 1983.

\bibitem{deGeresem2004finite}
H.~De~Gersem and T.~Weiland, ``Finite-element models for superconductive cables
  with finite interwire resistance,'' \emph{IEEE Transactions on Magnetics}, 
  vol.~40, no.~2, pp. 667--670, Mar. 2004.

\bibitem{verweij1995electrodynamics}
A.~P. Verweij, ``Electrodynamics of superconducting cables in accelerator
  magnets,'' Ph.D. dissertation, Universiteit Twente, 1995.

\bibitem{logan2012}
L.~D. Logan, \emph{A First Course in the Finite Element Method}.\hskip 1em plus
  0.5em minus 0.4em\relax Cengage Learning, 2012.

\bibitem{Savary2015}
F.~Savary, G.~Apollinari, B.~Auchmann, E.~Barzi, G.~Chlachidze, M.~Guinchard,
  P.~Grosclaude, S.~I. Bermudez, M.~Karppinen, C.~Löffler, G.~Kirby,
  C.~Kokkinos, F.~Lackner, T.~J. Lyon, A.~Nobrega, I.~Novitski, L.~Oberli,
  J.~C. Perez, F.~O. Pincot, L.~Rossi, J.~Rysti, G.~Willering, and A.~Zlobin,
  ``Design, assembly, and test of the CERN 2-m long 11 T dipole in single coil
  configuration,'' \emph{IEEE Transactions on Applied Superconductivity},
  vol.~25, no.~3, June 2015.

\end{thebibliography}
 \end{document}